# Vortex-bubble system as a spin acoustic particle model

Ion Simaciu[a,*], Gheorghe Dumitrescu[b], Zoltan Borsos[c,**], Viorel Drafta[d], Anca Baciu[c] and Georgeta Nan[c]

[a] Retired lecturer, Petroleum-Gas University of Ploiești, Ploiești 100680, Romania

[b] Retired professor, High School Toma N. Socolescu, Ploiești, Romania

[c] Petroleum-Gas University of Ploiești, Ploiești 100680, Romania

[d] Independent researcher

E-mail: [*] isimaciu@yahoo.com; [**] borzolh@upg-ploiesti.ro

Abstract:
A vortices-bubble system may be depicted by some features which may lead to infer its acoustic charge and its spin angular momentum. We study the dynamics of this system within the framework of fluid physics and with the help of Maxwell's hydrodynamic equations. Since we will adopt an approach of a fluid without viscosity and without gravity, then the fluid is in steady state if also we will neglect the acoustic radiation. When we expressed the energy (the kinetic and potential energy) of the vortex-bubble system we found out that the state of bubble which is captured in the core of the vortex is more stable than it is out of the core. For the above allegations we consider that this system, namely a vortex-bubble system, is a good approach for an acoustic particle with charge and spin angular momentum. This system owns to the Acoustic World.

## 1. Introduction

In this paper we study the system consisting of two components specific to the Acoustic World (AW), the pulsating bubble and the core vortex [1-5]. The need to study this system was due to the fact that the two components of AW have complementary properties specific to a charged acoustic particle model and having internal angular momentum. The pulsating bubble has the properties of an acoustic monopole, characterized by the acoustic charge [1, 2]. Two pulsating bubbles interact through secondary Bjerknes forces analogous to electrostatic forces [1, 2]. The vortex is the specific AW component that has the properties of an acoustic dipole. The forces between two vortices are proportional to $1/r^3$, $r > D$ [3-5]. The study of the bubble-vortex system is necessary by analogy with the classical and quantum electron models [6-8]. These models attempt to reproduce the experimental properties of the electron as an elementary particle. The electron is a particle charged with an electric charge and which also has a magnetic moment and therefore an internal angular momentum. Experiments prove that the electric charge is dispersed in a volume of the order $r_{exp} \cong 10^{-19}$ m [9]. Experiments also highlight the possibility of separating the three properties: electric charge, orbital angular momentum and spin angular momentum into: holon, orbiton and spinon [10-12]. When electron is not reduced to a point then the center of mass (CM) does not coincide with the center of the charge (CC), charge being a point, and then CC has an orbital motion around the center of mass. The radius of this motion is of the order of the Compton length $\lambda_C = h/(m_e c) \cong 10^{-13}$ m [6-8, 13].

In recent papers [14], the vortex structure of the electron as an extended particle is highlighted or the vortex properties of electron beams are highlighted [15, 16].

In the second part of the paper we study the dynamics of the vortex-bubble system in order to find out the connection between the system parameters under conditions where the system is stable, i.e. it has circular orbits. We find the acceleration of the bubble in the vortex field and the acceleration of the vortex in the bubble field using the results of the paper "On the motion of small spherical bubbles in two-dimensional vortical flows", in Cartesian coordinates and polar coordinates [17]. In the third part of the paper we study the dynamics of the system using Maxwell's hydrodynamic equations. In the fourth part of the paper we calculate the total kinetic momentum of the system and the total energy. The fifth part consists of discussions and conclusions based on the obtained results.

## 2. Dynamic modeling of the vortex–bubble system

### 2.1. Dynamics of the pulsating bubble in the vortex field

In the specialized scientific literature, the interaction between a vortex and a rigid spherical body or a bubble was studied [17-19]. Let be a pulsating bubble moving under the action of a vortex. The axis of the vortex passes through the center of the reference frame, $Oxyz$. The center of the bubble is placed in the plane perpendicular to the vortex axis and has the position vector $\vec{r} = \vec{r}_b$, relative to the vortex axis. Considering a Rankin-type vortex [20] in a fluid of density $\rho$ and viscosity coefficient zero, $\mu = 0$, with the core with radius $D$, the velocity vector of the vortex is: a) for the core of the vortex (the rigid-body vortex with the angular velocity, $\vec{\Omega}$)

$$v_{vxi} = -\Omega y, v_{vyi} = \Omega x, \vec{v}_{vi} = \Omega r \hat{\theta}, \quad r \leq D \quad (1)$$

and b) for the outside of the vortex core (the irrotational vortex)

$$v_{vxe} = \frac{-\Omega D^2 y}{r^2}, v_{vye} = \frac{\Omega D^2 x}{r^2}, \vec{v}_{ve} = \frac{\Omega D^2 r \hat{\theta}}{r}, \quad r > D. \quad (2)$$

We study the movement of the two interacting systems considering them independent: we study the movement of the pulsating bubble in the field of the vortex and we study the movement of the vortex in the field of the pulsating bubble. According to Eq. (4) from the paper [17], the equations of the dynamics of the bubble having the volume $V$ and density $\rho_b$, written on the components, in the presence of a vortex in a non-viscous fluid and in the absence of the gravitational interaction ($\vec{g} = 0$), are:

$$\dot{\vec{v}} = \frac{2\rho}{2\rho_b + \rho}\left(\frac{D}{Dt} + \frac{d}{2dt}\right)\vec{u},$$

$$\dot{v}_x = \frac{2\rho}{2\rho_b + \rho}\left(\frac{Du_x}{Dt} + \frac{du_x}{2dt}\right), \dot{v}_y = \frac{2\rho}{2\rho_b + \rho}\left(\frac{Du_y}{Dt} + \frac{du_y}{2dt}\right), \quad (3)$$

whth $\vec{r} = \vec{r}_b$, $\vec{v} = \vec{v}_b$, $\vec{u} = \vec{v}_v$ and the derivative following a fluid element and derivative along the particle path:

$$\frac{D}{Dt} = \frac{\partial}{\partial t} + (\vec{u} \cdot \nabla), \quad \frac{d}{dt} = \frac{\partial}{\partial t} + (\vec{v} \cdot \nabla). \quad (4)$$

In living Eqs. (4) in Eqs. (3), by components, results:

$$\dot{\upsilon}_{bx} = \frac{2\rho}{2\rho_b + \rho}\left[\frac{3}{2}\frac{\partial u_x}{\partial t} + \left(u_x + \frac{\upsilon_{bx}}{2}\right)\frac{\partial u_x}{\partial x} + \left(u_y + \frac{\upsilon_{by}}{2}\right)\frac{\partial u_x}{\partial y}\right]_{\vec{r}=\vec{r}_b},$$

$$\dot{\upsilon}_{by} = \frac{2\rho}{2\rho_b + \rho}\left[\frac{3}{2}\frac{\partial u_y}{\partial t} + \left(u_x + \frac{\upsilon_{bx}}{2}\right)\frac{\partial u_y}{\partial x} + \left(u_y + \frac{\upsilon_{by}}{2}\right)\frac{\partial u_y}{\partial y}\right]_{\vec{r}=\vec{r}_b}.$$

(5)

If we substitute the vortex-driven components of the fluid velocities (1) and (2) into Eqs. (5): a) in the vortex core, $u_{xi} = -\Omega y, u_{yi} = \Omega x$ and b) outside the vortex core $u_{xe} = -\Omega D^2 y/r^2$, $u_{ye} = \Omega D^2 x/r^2$ with $\partial u_x/\partial t = 0$, $\partial u_y/\partial t = 0$, we obtain: a)

$$\dot{\upsilon}_{bxi} = \frac{-2\rho\Omega}{2\rho_b + \rho}\left(\Omega x_b + \frac{\upsilon_{byi}}{2}\right), \dot{\upsilon}_{byi} = \frac{-2\rho\Omega}{2\rho_b + \rho}\left(\Omega y_b - \frac{\upsilon_{bxi}}{2}\right), r_b \leq D \quad (6)$$

or

$$\dot{\vec{\upsilon}}_{bi} = \frac{-2\rho\Omega}{2\rho_b + \rho}\left[\Omega(x_b\hat{x} + y_b\hat{y}) + \frac{(\upsilon_{byi}\hat{x} - \upsilon_{bxi}\hat{y})}{2}\right] =$$

$$-A\Omega\left[\Omega(x_b\hat{x} + y_b\hat{y}) + \frac{(\upsilon_{byi}\hat{x} - \upsilon_{bxi}\hat{y})}{2}\right], A = \frac{2\rho}{2\rho_b + \rho}, r_b \leq D \quad (7)$$

and b)

$$\dot{\upsilon}_{bxe} = \frac{-2\rho\Omega D^2}{(2\rho_b + \rho)r_b^2}\left(\frac{\Omega D^2 x_b}{r_b^2} + \frac{\upsilon_{bye}}{2}\right), \dot{\upsilon}_{bye} = \frac{-2\rho\Omega D^2}{(2\rho_b + \rho)r_b^2}\left(\frac{\Omega D^2 y_b}{r_b^2} - \frac{\upsilon_{bxe}}{2}\right), r_b > D \quad (8)$$

or

$$\dot{\vec{\upsilon}}_{be} = \frac{-2\rho\Omega D^2}{(2\rho_b + \rho)r_b^2}\left[\frac{\Omega D^2(x_b\hat{x} + y_b\hat{y})}{r_b^2} + \frac{(\upsilon_{bye}\hat{x} - \upsilon_{bxe}\hat{y})}{2}\right] =$$

$$\frac{-A\Omega D^2}{(2\rho_b + \rho)r_b^2}\left[\frac{\Omega D^2(x_b\hat{x} + y_b\hat{y})}{r_b^2} + \frac{(\upsilon_{bye}\hat{x} - \upsilon_{bxe}\hat{y})}{2}\right], r_b > D. \quad (9)$$

In order to find the expressions of the accelerations of the bubble (more generally, of the spherical body with density $\rho_b$) in the vortex field, we make the hypothesis that the speed of the bubble is proportional to the speed of the vortex, the proportionality constant being a positive real number, $B > 0$ : a) in the core of the vortex

$$\upsilon_{\upsilon xi} = -B\Omega y, \upsilon_{\upsilon yi} = B\Omega x, \vec{\upsilon}_{\upsilon i} = B\Omega r\hat{\theta}, r \leq D \quad (10)$$

and b) outside the vortex core

$$\upsilon_{\upsilon xe} = \frac{-B\Omega D^2 y}{r^2}, \upsilon_{\upsilon ye} = \frac{-B\Omega D^2 x}{r^2}, \vec{\upsilon}_{\upsilon e} = \frac{B\Omega D^2\hat{\theta}}{r}, r > D. \quad (11)$$

Substituting these velocity expressions into Eqs. (7) and (9), it results: a) in the core of the vortex

$$\dot{\vec{\upsilon}}_{bi} = -A\Omega\left[\Omega(x_b\hat{x} + y_b\hat{y}) + \frac{B\Omega(x_b\hat{x} + y_b\hat{y})}{2}\right] = -A\left(1 + \frac{B}{2}\right)\Omega^2\vec{r}_b, r_b \leq D \quad (12)$$

and b) in the outer core vortex

$$\vec{v}_{be} = -A\left(1+\frac{B}{2}\right)\Omega^2 D^4 \frac{\vec{r}_b}{r_b^4}, \quad r_b > D. \tag{13}$$

We find the constant $B > 0$ from the condition that the centripetal acceleration of the bubble $\left|\dot{\vec{v}}_b\right| = a_{cp}$ has the same modulus as the centrifugal acceleration $a_{cf} = \vec{v}^2/r_b$ i.e. the circular trajectory condition (the condition of the bubble's stable orbit in the vortex field)

$$\left|\dot{\vec{v}}_b\right| = a_{cp} = a_{cf} = \frac{\vec{v}_b^2}{r_b}. \tag{14}$$

Substituting into equality (14) the expressions given by Eqs. (10-13), it follows: a)

$$A\Omega^2 r_b\left(1+\frac{B}{2}\right) = B^2\Omega^2 r_b \text{ sau } B^2 - \frac{AB}{2} - A = 0, \quad r_b \leq D \tag{15}$$

and b)

$$\frac{A\Omega^2 D^4}{r_b^3}\left(1+\frac{B}{2}\right) = \frac{B^2\Omega^2 D^4}{r_b^3} \text{ sau } B^2 - \frac{AB}{2} - A = 0, \quad r_b > D, \tag{16}$$

that is, the same equation with the unknown the constant $B > 0$. The solutions of Eqs. (15, 16) are

$$B_{1,2} = \frac{A}{4}\left(1 \pm \sqrt{1+\frac{16}{A}}\right). \tag{17}$$

The real positive solution is $B = (A/4)(1+\sqrt{1+16/A})$. Substituted into the bubble velocity and acceleration expressions, Eqs. (10-13), we obtain the expressions: a), $r_b \leq D$,

$$\vec{v}_{bi} = \frac{A}{4}\left(1+\sqrt{1+\frac{16}{A}}\right)\Omega r_b \hat{\theta} = \frac{\left(1+3\sqrt{1+16\rho_b/(9\rho)}\right)\Omega r_b \hat{\theta}}{2(1+2\rho_b/\rho)}, \tag{18}$$

$$\dot{\vec{v}}_{bi} = -A\Omega^2\left[1+\frac{A}{8}\left(1+\sqrt{1+\frac{16}{A}}\right)\right]\vec{r}_b = \frac{-2\Omega^2}{(1+2\rho_b/\rho)}\left[1+\frac{\left(1+3\sqrt{1+16\rho_b/(9\rho)}\right)}{4(1+2\rho_b/\rho)}\right]\vec{r}_b = -C\Omega^2\vec{r}_b, \tag{19}$$

with the constant $(A/4)^2(1+\sqrt{1+16/A})^2 = C$, and b), $r_b > D$,

$$\vec{v}_{be} = \frac{\frac{A}{4}\left(1+\sqrt{1+\frac{16}{A}}\right)\Omega D^2 \hat{\theta}}{r_b} = \frac{\left(1+3\sqrt{1+16\rho_b/(9\rho)}\right)\Omega D^2 \hat{\theta}}{2(1+2\rho_b/\rho)r_b}, \tag{20}$$

$$\dot{\vec{v}}_{be} = \frac{-A\Omega^2 D^4}{r_b^4}\left[1+\frac{A}{8}\left(1+\sqrt{1+\frac{16}{A}}\right)\right]\vec{r}_b = \frac{-2\rho\Omega^2 D^4}{(2\rho_b+\rho)r_b^4}\left[1+\frac{\left(1+3\sqrt{1+16\rho_b/(9\rho)}\right)}{4(1+2\rho_b/\rho)}\right]\vec{r}_b =$$

$$= \frac{-C\Omega^2 D^4}{r_b^4}\vec{r}_b, \quad C = \frac{2}{(1+2\rho_b/\rho)}\left[1+\frac{\left(1+3\sqrt{1+16\rho_b/(9\rho)}\right)}{4(1+2\rho_b/\rho)}\right]. \tag{21}$$

The solutions of the velocities given by Eqs. (18, 20) differs from the expression of the speed of a spherical body driven into motion by the moving fluid (($\vec{v}_{bi} = 3\rho\vec{v}_{vi}/(2\rho_b+\rho)$, $r \leq D$; $\vec{v}_{be} = 3\rho\vec{v}_{ve}/(2\rho_b+\rho)r$, $r > D$); deduced in the book [21 - Sch. 11].

As a check of the correctness of the expressions (18-21), in the situation when the density of the bubble is equal to the density of the fluid, $\rho_b = \rho$, the speed and acceleration of the spherical body should be equal to the speed and acceleration of the vortex, at the same radius $r_b$, because the constant is $A = 2/3$. Substituting the condition of equality of densities into Eqs. (18-21), it follows: a)

$$\vec{v}_{bi} = \vec{v}_{vi} = \Omega r_b \hat{\theta}, \quad \dot{\vec{v}}_{bi} = \dot{\vec{v}}_{vi} = -\Omega^2 \vec{r}_b, \quad r_b \leq D \qquad (22)$$

and b)

$$\vec{v}_{be} = \vec{v}_{ve} = \frac{\Omega D^2 \hat{\theta}}{r_b}, \quad \dot{\vec{v}}_{be} = \dot{\vec{v}}_{ve} = \frac{-\Omega^2 D^4 \vec{r}_b}{r_b^4}, \quad r_b > D. \qquad (23)$$

For the situation when $\rho_b \ll \rho$, i.e. $A \cong 2$, expressions (18-21), can be approximated: a)

$$\vec{v}_{bi} = \frac{A}{4}\left(1 + \sqrt{1 + \frac{16}{A}}\right)\Omega r_b \hat{\theta} \cong \frac{2\rho \Omega r_b \hat{\theta}}{(2\rho_b + \rho)} \cong 2\Omega r_b \hat{\theta}, \quad r_b \leq D; \qquad (24)$$

$$\dot{\vec{v}}_i = -A\Omega^2\left[1 + \frac{A}{8}\left(1 + \sqrt{1 + \frac{16}{A}}\right)\right]\vec{r}_b \cong -\left(\frac{2\rho \Omega^2}{2\rho_b + \rho}\right)\vec{r}_b = -4\Omega^2 \vec{r}_b, \quad r_b \leq D \qquad (25)$$

and b)

$$\vec{v}_{ve} \cong \frac{A\Omega D^2 \hat{\theta}}{r_b} = \frac{2\rho \Omega D^2 \hat{\theta}}{(2\rho_b + \rho)r_b} \cong \frac{2\Omega D^2 \hat{\theta}}{r_b}, \quad r_b > D. \qquad (26)$$

$$\dot{\vec{v}}_e \cong \frac{-2A\Omega^2 D^4 \vec{r}_b}{r_b^4} = -\left(\frac{2\rho}{2\rho_b + \rho}\right)\frac{2\Omega^2 D^4 \vec{r}_b}{r_b^4} \cong -\frac{4\Omega^2 D^4 \vec{r}_b}{r_b^4}, \quad r_b > D. \qquad (27)$$

According to the expressions of the accelerations given by Eqs. (19, 21) and the stability condition (14), the bubble performs a circular motion on orbits of radius $r_b$, if we neglect the radial movement of the vortex center under the action of the acoustic field generated by the bubble pulsation.

We get the same result using polar coordinates ($r, \theta$). According to Appendix D from the work Elementary Fluid Mechanics [4], the convection term $(\vec{u} \cdot \nabla)\vec{u}$ in polar coordinates ( $\vec{u} = u_r \hat{r} + u_\theta \hat{\theta}$ ) is

$$(\vec{u} \cdot \nabla)\vec{u} = \left(u_r \frac{\partial u_r}{\partial r} + \frac{u_\theta}{r}\frac{\partial u_r}{\partial \theta} - \frac{u_\theta^2}{r}\right)\hat{r} + \left(u_r \frac{\partial u_\theta}{\partial r} + \frac{u_\theta}{r}\frac{\partial u_\theta}{\partial \theta}\right)\hat{\theta} \qquad (28)$$

and the term $(\vec{v} \cdot \nabla)\vec{u}$ is

$$(\vec{v} \cdot \nabla)\vec{u} = \left(v_r \frac{\partial u_r}{\partial r} + \frac{v_\theta}{r}\frac{\partial u_r}{\partial \theta} - \frac{v_\theta u_\theta}{r}\right)\hat{r} + \left(v_r \frac{\partial u_\theta}{\partial r} + \frac{v_\theta}{r}\frac{\partial u_\theta}{\partial \theta}\right)\hat{\theta} \qquad (29)$$

Substituting Eqs. (28, 29) in the derivative following a fluid element and derivative along the particle path, results:

$$\frac{D\vec{u}}{Dt} = \frac{\partial \vec{u}}{\partial t} + (\vec{u} \cdot \nabla)\vec{u} = \left(\frac{\partial u_r}{\partial t} + u_r \frac{\partial u_r}{\partial r} + \frac{u_\theta}{r}\frac{\partial u_r}{\partial \theta} - \frac{u_\theta^2}{r}\right)\hat{r} + \left(\frac{\partial u_\theta}{\partial t} + u_r \frac{\partial u_\theta}{\partial r} + \frac{u_\theta}{r}\frac{\partial u_\theta}{\partial \theta}\right)\hat{\theta}, \qquad (30)$$

$$\frac{d\vec{u}}{dt} = \frac{\partial \vec{u}}{\partial t} + (\vec{v} \cdot \nabla)\vec{u} = \left(\frac{\partial u_r}{\partial t} + v_r \frac{\partial u_r}{\partial r} + \frac{v_\theta}{r}\frac{\partial u_r}{\partial \theta} - \frac{v_\theta u_\theta}{r}\right)\hat{r} + \left(\frac{\partial u_\theta}{\partial t} + v_r \frac{\partial u_\theta}{\partial r} + \frac{v_\theta}{r}\frac{\partial u_\theta}{\partial \theta}\right)\hat{\theta}. \qquad (31)$$

With these expressions, the acceleration of the vortex-entrained bubble is

$$\dot{\vec{v}}_b = \left(\frac{D\vec{u}}{Dt} + \frac{d\vec{u}}{2dt}\right) = \left\{\hat{r}\left[\frac{3\partial u_r}{2\partial t} + \left(u_r + \frac{v_r}{2}\right)\frac{\partial u_r}{\partial r} + \left(\frac{u_\theta}{r} + \frac{v_\theta}{2r}\right)\frac{\partial u_r}{\partial \theta} - \left(\frac{u_\theta^2}{r} + \frac{v_\theta u_\theta}{2r}\right)\right]_{\vec{r}=\vec{r}_b} + \right.$$
$$\left. \hat{\theta}\left[\frac{3\partial u_\theta}{2\partial t} + \left(u_r + \frac{v_r}{2}\right)\frac{\partial u_\theta}{\partial r} + \left(\frac{u_\theta}{r} + \frac{v_\theta}{2r}\right)\frac{\partial u_\theta}{\partial \theta}\right]_{\vec{r}=\vec{r}_b}\right\}, \quad A = \frac{2\rho}{2\rho_b + \rho}. \quad (32)$$

The components of bubble acceleration are:

$$\dot{v}_{br} = A\left[\frac{3\partial u_r}{2\partial t} + \left(u_r + \frac{v_{br}}{2}\right)\frac{\partial u_r}{\partial r} + \left(\frac{u_\theta}{r} + \frac{v_{b\theta}}{2r}\right)\frac{\partial u_r}{\partial \theta} - \left(\frac{u_\theta^2}{r} + \frac{v_{b\theta}u_\theta}{2r}\right)\right]_{\vec{r}=\vec{r}_b},$$
$$\dot{v}_{b\theta} = A\left[\frac{3\partial u_\theta}{2\partial t} + \left(u_r + \frac{v_{br}}{2}\right)\frac{\partial u_\theta}{\partial r} + \left(\frac{u_\theta}{r} + \frac{v_{b\theta}}{2r}\right)\frac{\partial u_\theta}{\partial \theta}\right]_{\vec{r}=\vec{r}_b}. \quad (33)$$

Substituting the core fluid velocity components $u_r = \dot{r} = 0, u_\theta = \dot{\theta}r = \Omega r$ into Eq. (33), it follows:

$$\dot{v}_{br} = -A\Omega^2 r_b\left(1 + \frac{v_{b\theta}}{2\Omega r_b}\right), \quad \dot{v}_{b\theta} = A\left(\frac{v_{br}\Omega}{2}\right). \quad (34)$$

The minus sign for the radial acceleration of the bubble means that it is a centripetal acceleration. From the condition that the motion of the bubble is circular (that is $r_b = const.$ or $v_{br} = 0$) and Eqs. (34), it follows:

$$\dot{v}_{b\theta} = 0, v_{b\theta} = B\Omega r_b, B > 0; \quad |\dot{v}_{br}| = A\Omega^2 r_b\left(1 + \frac{v_{b\theta}}{2\Omega r_b}\right) = A\Omega^2 r_b\left(1 + \frac{B}{2}\right). \quad (35)$$

For the dynamic stability of the bubble, it is necessary that the centripetal acceleration is equal to the centrifugal acceleration ($|\dot{v}_{br}| \equiv a_{cp} = a_{cf} \equiv v_{b\theta}^2/r_b$), according to Eq. (14),

$$A\Omega^2 r_b\left(1 + \frac{B}{2}\right) = B^2\Omega^2 r_b, \quad (36)$$

This relation is identical to Eq. (15) which represents the equation by which the constant $B$ is expressed as a function of the material constant $A$. From Eq. (36) yields the same solution given by Eq. (17) and therefore also in polar coordinates we find the same expressions of tangential velocity and radial acceleration given by Eqs. (18-21).

## 2.2. Vortex dynamics in the pulsating bubble field

Through a similar procedure, we also study the dynamics of the vortex, in the radial direction, under the action of the acoustic field generated by the pulsation of the bubble. According to Eq. (4) from the paper [17], the dynamics equations of the vortex having the volume $V$, in the presence of a pulsating bubble in a non-viscous fluid of density $\rho$ (în Eq. (5) $\rho_b \to \rho$) and in the absence of gravitational interaction ($\vec{g} = 0$), are:

$$\dot{\vec{v}}_{vr} = \frac{2}{3}\left(\frac{D}{Dt} + \frac{d}{dt}\right)\vec{u}_b. \quad (37)$$

Substituting Eqs. (3) in Eq. (37), it follows:

$$\dot{\vec{\upsilon}}_{\upsilon r} = \frac{2}{3}\left[\frac{3}{2}\frac{\partial \vec{u}_b}{\partial t} + \left(\left(u_x + \frac{\upsilon_{\upsilon rx}}{2}\right)\frac{\partial u_{bx}}{\partial x} + \left(u_y + \frac{\upsilon_{\upsilon ry}}{2}\right)\frac{\partial u_{bx}}{\partial y}\right)\bigg|_{\vec{r}=\vec{r}_\upsilon}\hat{x} + \right.$$
$$\left.\left(\left(u_x + \frac{\upsilon_{\upsilon rx}}{2}\right)\frac{\partial u_{by}}{\partial x} + \left(u_y + \frac{\upsilon_{\upsilon ry}}{2}\right)\frac{\partial u_{by}}{\partial y}\right)\bigg|_{\vec{r}=\vec{r}_\upsilon}\hat{y}\right]. \quad (38)$$

The bubble, with the equilibrium radius $R_0$, the pulsation $\omega$, the dimensionless amplitude of the pulsating motion $a \ll 1$ and the initial phase of the pulsation $\varphi_b$ [1], pulsates according to the relations:

$$R(t) = R_0\left[1 + a\cos(\omega t + \varphi_b)\right], \dot{R}(t) = -\omega a R_0 \sin(\omega t + \varphi_b). \quad (39)$$

We find the components of the fluid velocity generated by the pulsation of the bubble using the expression of the velocity potential [22]:

$$\Phi_b = \frac{-\dot{R}R_0^2}{\left[(x-x_2)^2 + (y-y_2)^2 + z^2\right]^{1/2}}, u_{bx} = \frac{\partial \Phi_b}{\partial x}\bigg|_{z=0} = \frac{-\dot{R}R_0^2 x}{(x^2+y^2)^{3/2}}, u_{by} = \frac{\partial \Phi_b}{\partial y}\bigg|_{z=0} = \frac{-\dot{R}R_0^2 y}{(x^2+y^2)^{3/2}}. \quad (40)$$

Using the relations given by Eq. (40), we also find the time variation of these velocities:

$$\frac{\partial u_{bx}}{\partial t} = \frac{-\ddot{R}R_0^2 x}{(x^2+y^2)^{3/2}}, \quad \frac{\partial u_{by}}{\partial t} = \frac{-\ddot{R}R_0^2 y}{(x^2+y^2)^{3/2}}. \quad (41)$$

Substituting into Eq. (38) the expressions of the velocity components, we obtain:

$$\dot{\vec{\upsilon}}_{\upsilon r} = \frac{\ddot{R}R_0^2 \vec{r}_\upsilon}{r_\upsilon^3} - \frac{4\dot{R}^2 R_0^4 \vec{r}_\upsilon}{3r_\upsilon^6} - \frac{2\dot{R}R_0^2 \vec{\upsilon}_{\upsilon r}}{3r_\upsilon^5}, \quad (42)$$

with $\vec{\upsilon}_{\upsilon r} = \upsilon_\upsilon \vec{r}_\upsilon / r_\upsilon$. For $r_\upsilon \gg R_0$ and $\ddot{R} \gg \dot{R}^2$, Eq. (42) is approximated by:

$$\frac{d\upsilon_\upsilon}{dt} \cong \frac{\ddot{R}R_0^2}{r_\upsilon^2} = \frac{R_0^2}{r_\upsilon^2}\frac{d\dot{R}}{dt} \text{ sau } d\upsilon_\upsilon = \frac{R_0^2}{r_\upsilon^2}d\dot{R}. \quad (43)$$

The solution of Eq. (43) is:

$$\upsilon_{\upsilon r} = \frac{R_0^2}{r_\upsilon^2}\left(\dot{R} - \dot{R}(t=0)\right). \quad (44)$$

Substituting the expression $\dot{R}(t)$ given by Eq. (39) in Eq. (44), it follows:

$$\vec{\upsilon}_{\upsilon r} = -\frac{R_0^3 a \omega \vec{r}_\upsilon}{r_\upsilon^3}\left(\sin(\omega t + \varphi_b) - \sin\varphi_b\right), \vec{r}_\upsilon = -\vec{r}_b. \quad (45)$$

If we consider the vortex axis fixed, the velocity given by Eqs. (44, 45) is the radial velocity of the bubble relative to the vortex. According to Eqs. (18, 20, 45), considering the center of the fixed vortex, the center of the bubble simultaneously executes a rotational movement in the same direction as the vortex and an oscillatory radial movement in the direction of the vector $\vec{r}_\upsilon = -\vec{r}_b$. With the help of Eq. (43) and the definition of the radial velocity of the vortex, $\upsilon_{\upsilon r} = dr_\upsilon/dt$, we can find the time variation of the vector modulus $\vec{r}_\upsilon(t)$ from the equation:

$$\frac{dr_\upsilon}{dt} = \frac{R_0^2}{r_\upsilon^2}\frac{dR}{dt} \text{ or } r_\upsilon^2 dr_\upsilon = R_0^2 dR, \quad (46)$$

which becomes integrated

$$r_v(t) = r_{v0}\left[1 + \frac{3R_0^3}{r_{v0}^3}\left(\cos(\omega t + \varphi_b) - \cos\varphi_b\right)\right]^{\frac{1}{3}} \cong r_{v0}\left[1 + \frac{R_0^3}{r_{v0}^3}\left(\cos(\omega t + \varphi_b) - \cos\varphi_b\right)\right], \quad (47)$$

with $r_v(t=0) = r_{v0}$. The result obtained, Eq. (47), highlights the relative oscillatory motion of the vortex-bubble system in the radial direction. In this study of the motion of the vortex–bubble system we have neglected the motion of the vortex under the action of the plane wave that maintains the pulsating motion of the bubble.

## 3. Modeling of the vortex-bubble system with Maxwell's hydrodynamic equations

The modeling of the vortex-bubble system can also be solved with the help of Maxwell's hydrodynamic equations [2, 23]. According to the paper [2], the pulsating bubble has the properties of an acoustic monopole which is characterized by the acoustic charge:

$$q_b = V\rho \quad (48)$$

and through the acoustic intensity of the field being around the bubble

$$\vec{E}_b \cong \frac{\ddot{V}\hat{r}}{4\pi r^2}. \quad (49)$$

According to the paper [3], the vortex has the properties of an acoustic dipole which is characterized by the acoustic field vectors: a) inside the core, $\Omega^2 r_b = \vec{F}_{vi}/(V\rho)$,

$$\vec{E}_{vi} = \left(-\Omega^2 r\right)\hat{r}, \quad \vec{H}_{vi} = 2\vec{\Omega} = 2\Omega\hat{z} \quad (50)$$

and b) outside the core, $\Omega^2 D^4/r^3 = \vec{F}_{ve}/(V\rho)$,

$$\vec{E}_{ve} = \frac{\Omega^2 D^4}{r^4}\vec{r}. \quad (51)$$

With these physical quantities, the force exerted by the vortex on the bubble is: a) inside the core

$$\vec{F}_{vi} \equiv q_b\left[\vec{E}_{vi} + \left(\vec{v}_{bi} \times \vec{H}_{vi}\right)\right] = \rho V\left[\left(-\Omega^2 r_b\right)\hat{r} + \left(\Omega r_b\hat{\theta} \times 2\Omega\hat{z}\right)\right] = \rho V\Omega^2 r_b\left(-\hat{r} + 2\hat{r}\right) = \rho V\Omega^2 r_b\hat{r} \quad (52)$$

and b) outside the core

$$\vec{F}_{ve} \equiv q_b\vec{E}_{ve} = \left(\frac{\rho V\Omega^2 D^4}{r_b^3}\right)\hat{r}. \quad (53)$$

The radial accelerations corresponding to these forces are: a) $\Omega^2 r_b = \vec{F}_{vi}/(V\rho)$, inside the core and b) $\Omega^2 D^4/r^3 = \vec{F}_{ve}/(V\rho)$, outside the core. They differ from the accelerations obtained in the second section by the constant that depends on the densities of the fluid and the spherical body (bubble) and by the sign (they are centrifugal, i.e. oriented in the direction of the vector $\vec{r}_b$). This difference between the two methods highlights a mistake in Maxwell's hydrodynamic equations corresponding to the two systems specific to the acoustic world.

The forces exerted by the bubble, with the intensity of the acoustic field given by Eq. (49), on the vortex that has the charge per unit length [3]

$$q_v = e_v\varepsilon_v = \left(-\Omega^2 D^2\right)\frac{\rho}{\Omega^2} = -D^2\rho \quad (54)$$

it is

$$\frac{\vec{F}_{b\upsilon}}{l} \equiv q_\upsilon \vec{E}_b = \left(-D^2\rho\right)\frac{\ddot{V}\hat{r}}{4\pi r_\upsilon^2} = \frac{-V\rho}{l}\frac{\ddot{V}\hat{r}}{4\pi r_\upsilon^2}, \vec{F}_{b\upsilon} = \frac{(-V\rho)\ddot{V}\hat{r}}{4\pi r_\upsilon^2}, r_\upsilon = r_b, \tag{55}$$

for both situations i.e. the vortex inside and the vortex outside the core.

The acceleration of the vortex under the action of the acoustic field of the bubble is an oscillatory acceleration $-\ddot{V}\hat{r}/(4\pi r_\upsilon^2) \equiv \vec{F}_{b\upsilon}/(V\rho) = -\ddot{R}R_0^2\hat{r}/r_\upsilon^2$ like the acceleration given by Eq. (43), minus the sign.

According to the results obtained in sections 2 and 3, it follows that the vortex-bubble system is a stable system, the bubble performing simultaneously a circular motion (the radius of the orbit is $r_\upsilon(t) = r_b(t)$) and a small-amplitude oscillating radial motion of the order of $R_0^3/r_{\upsilon 0}^2$, according to Eq. (47). The formed system simultaneously has a monopolar acoustic charge determined by the charge of the oscillating bubble, an acoustic charge determined by the vortex, and an intrinsic kinetic momentum determined by the sum of the kinetic momentum of the vortex and the orbital momentum of the bubble.

## 4. Specific properties of the vortex-bubble system

### 4.1. Angular momentum of the system

The angular momentum of the vortex-bubble system is obtained between the angular momentum of the vortex and the orbital momentum of the bubble, because, as vectors, they have the same orientation given by the angular velocity vector $\vec{\Omega} = \Omega\hat{z}$:

$$L_{b\upsilon} = S_\upsilon + L_{bo}, \tag{56}$$

The angular momentum of the vortex was calculated in the paper Vortexes as systems specific to the Acoustic World [4]. This angular moment depends on the radius $R_{en}$ of the enclosure in which the bubble-vortex system exists, on the height of the vortex $l$ in the direction perpendicular to the plane $Oxy$ and is given, according to Eq. 41 of [4] of the expression

$$S_\upsilon = \int_0^{R_{en}} dm_\upsilon \upsilon_\upsilon r = \int_0^{R_{en}} (2\pi\rho rl dr)\upsilon_\upsilon r = \pi\rho l D^2 R_{en}^2 \Omega\left(1 - \frac{D^2}{2R_{en}^2}\right). \tag{57}$$

The orbital kinetic momentum of the bubble with virtual mass, $m_b = 2\pi R^3 \rho/3$ [1], which is driven by the vortex, is given by the expression

$$L_{bo} = m_b \upsilon_{b\upsilon} r_b = \frac{2\pi\rho R^3 r_b}{3}\upsilon_{b\upsilon}. \tag{58}$$

Substituting the bubble velocity expressions given by Eqs. (18, 20), in Eq. (58), we obtain: a) for the core of the vortex

$$L_{boi} = \frac{\left(1+3\sqrt{1+16\rho_b/(9\rho)}\right)(\pi\rho R^3 r_b^2 \Omega)}{3(1+2\rho_b/\rho)}, \tag{59}$$

i.e. a angular momentum that depends on the radius of the bubble orbit and b) for the outer vortex

$$L_{boe} = \frac{\left(1+3\sqrt{1+16\rho_b/(9\rho)}\right)(\pi\rho R^3 D^2 \Omega)}{3(1+2\rho_b/\rho)}, \tag{60}$$

that is, an angular momentum that does not depend on the radius of the bubble's orbit.

Since the radius of the bubble and the radius of the orbit depend periodically on time: a) the orbital momentum inside the vortex $L_{boi}$ has an average value given by the relation:

$$\langle L_{boi} \rangle_t = \frac{\left(1+3\sqrt{1+16\rho_b/(9\rho)}\right)(\pi\rho\Omega)\langle R^3 \rangle_t \langle r_b^2 \rangle_t}{3(1+2\rho_b/\rho)} \cong \frac{\left(1+3\sqrt{1+16\rho_b/(9\rho)}\right)(\pi\rho R_0^3 r_{b0}^2 \Omega)}{3(1+2\rho_b/\rho)}, \quad (61)$$

with the orbit radius given by Eq. (47), $r_v(t=0) = r_{v0} = r_{b0}$ and b) the orbital momentum outside the vortex

$$\langle L_{boe} \rangle_t = \frac{\left(1+3\sqrt{1+16\rho_b/(9\rho)}\right)(\pi\rho D^2 \Omega)\langle R^3 \rangle_t}{3(1+2\rho_b/\rho)} \cong \frac{\left(1+3\sqrt{1+16\rho_b/(9\rho)}\right)(\pi\rho R_0^3 D^2 \Omega)}{3(1+2\rho_b/\rho)}, \quad (62)$$

Substituting the expressions given by Eqs. (59, 61) in Eq. (56), results: a) for the core of the vortex

$$\langle L_{bvi} \rangle_t = S_v + \langle L_{boi} \rangle_t \cong \pi\rho l D^2 R_{en}^2 \Omega \left(1 - \frac{D^2}{2R_{en}^2}\right) + \frac{\left(1+3\sqrt{1+16\rho_b/(9\rho)}\right)}{(1+2\rho_b/\rho)} \frac{\pi\rho R_0^3 r_{b0}^2 \Omega}{3} \quad (63)$$

and b) outside of the core

$$\langle L_{bve} \rangle_t = S_v + \langle L_{boe} \rangle_t \cong \pi\rho l D^2 R_{en}^2 \Omega \left(1 - \frac{D^2}{2R_{en}^2}\right) + \frac{\left(1+3\sqrt{1+16\rho_b/(9\rho)}\right)(\pi\rho R_0^3 D^2 \Omega)}{3(1+2\rho_b/\rho)}. \quad (64)$$

In the conditions $\rho \gg \rho_b$ and $R_{en} \gg D$, the expressions of the total angular moment is approximated by:

$$\langle L_{bvi} \rangle_t \cong \pi\rho l D^2 R_{en}^2 \Omega + \frac{4}{3}\pi\rho R_0^3 r_{b0}^2 \Omega = \pi\rho\left(lD^2 R_{en}^2 + \frac{4R_0^3 r_{b0}^2}{3}\right)\Omega,$$

$$\langle L_{bve} \rangle_t \cong \pi\rho l D^2 R_{en}^2 \Omega + \frac{4\pi\rho R_0^3 D^2 \Omega}{3} = \pi\rho D^2 \left(lR_{en}^2 + \frac{4R_0^3}{3}\right)\Omega, \quad (65)$$

that is, the total angular momentum of the vortex-bubble system connects enclosure parameters $R_{en}, \rho$ with vortex and bubble parameters $D, \Omega, l, R_0, r_{b0}$.

## 4.2. The energy of the vortex-bubble system

The kinetic energy of the system is obtained by summing of the kinetic energy of the vortex $E_{kv}$, the orbital kinetic energy of the bubble $E_{kbo}$, the energy of radial oscillation (considering the vortex axis fixed) $E_{kbvr}$ and the pulsation kinetic energy of the bubble $E_{kbp}$,

$$E_{kbv} = E_{kv} + E_{kbo} + E_{kbvr} + E_{kbp}. \quad (66)$$

We calculate the kinetic energy of the vortex according to the relation

$$E_{kv} = \int_0^{R_{en}} \frac{dm_v v_v^2}{2} = \int_0^D (\pi\rho r l dr)(\Omega r)^2 + \int_D^{R_{en}} (\pi\rho r l dr)\left(\frac{\Omega D^2}{r}\right)^2 =$$

$$\pi\rho l D^4 \Omega^2 \left(\frac{1}{4} + \ln\frac{R_{en}}{D}\right). \quad (67)$$

The kinetic energy of the bubble in orbital motion is, using the bubble velocity expressions given by Eqs. (18, 20): a) in the core of the vortex

$$E_{kboi} = \frac{m_b v_{bi}^2}{2} = \frac{m_b C \Omega^2 r_b^2}{2} = \frac{\pi\rho R^3 C \Omega^2 r_b^2}{3}, \quad r_b \leq D \quad (68)$$

and b) in the outer vortex

$$E_{kboe} = \frac{m_b \upsilon_{be}^2}{2} = \frac{m_b C D^4 \Omega^2}{2 r_b^2} \cong \frac{\pi \rho R^3 C D^4 \Omega^2}{3 r_b^2}, \quad r_b > D. \tag{69}$$

The kinetic energy of the bubble in radial motion is, using the expression for the radial velocity of the vortex (we consider the vortex axis fixed) given by Eqs. (44, 45):

$$E_{kbvr} = \frac{m_b \upsilon_{vr}^2}{2} = \frac{\pi \rho R^3 R_0^4 \left( \dot{R} - \dot{R}(t=0) \right)^2}{3 r_b^4}, \quad r_v = r_b. \tag{70}$$

The kinetic energy of the pulsating bubble is

$$E_{kbp} = \int_R^\infty \frac{dm_v \upsilon_p^2}{2} = \int_R^\infty \frac{4\pi r^2 \rho \, dr}{2} \left( \frac{\dot{R} R_0^2}{r^2} \right)^2 = 2\pi \rho \dot{R}^2 R_0^4 \int_R^\infty \frac{dr}{r^2} = \frac{2\pi \rho \dot{R}^2 R_0^4}{R} \cong 2\pi \rho R_0^3 \dot{R}^2. \tag{71}$$

The time-averaged orbital kinetic energies are: a) in the core of the vortex

$$\left\langle E_{kboi} \right\rangle_t = \frac{\pi \rho C \Omega^2 \left\langle R^3 r_b^2 \right\rangle_t}{3} \cong \frac{\pi \rho C \Omega^2 R_0^3 r_{b0}^2}{3} = \frac{m_{b0} C \Omega^2 r_{b0}^2}{2}, \quad r_b \leq D \tag{72}$$

and b) outside of the core

$$\left\langle E_{kboe} \right\rangle_t = \frac{\pi \rho C D^4 \Omega^2}{3} \left\langle \frac{R^3}{r_b^2} \right\rangle_t \cong \frac{\pi \rho R_0^3 C D^4 \Omega^2}{3 r_{b0}^2} = \frac{m_{b0} C D^4 \Omega^2}{2 r_{b0}^2}, \quad r_b > D. \tag{73}$$

The average kinetic energy of the bubble in pulsating motion is

$$\left\langle E_{kbp} \right\rangle_t = \left\langle 2\pi \rho \dot{R}^2 R_0^3 \right\rangle_t \cong 2\pi \rho R_0^5 a^2 \omega^2. \tag{74}$$

The averaged kinetic energy of the bubble in radial motion, according to averaging Eq. (70), is

$$\left\langle E_{kbvr} \right\rangle_t = \frac{\pi \rho R_0^4}{3} \left\langle \frac{R^3 \left( \dot{R} - \dot{R}(t=0) \right)^2}{r_b^4} \right\rangle_t \cong \frac{\pi \rho R_0^9 \omega^2 a^2 \sin^2 \varphi_b}{3 r_{b0}^4}, \quad r_{v0} = r_{b0} > R_0. \tag{75}$$

and is much smaller than the energy given by Eq. (71) of the bubble in pulsating motion.

The average total kinetic energy of the system is: a) in the core of the vortex

$$\left\langle E_{kbvi} \right\rangle_t = E_{kvi} + \left\langle E_{kboi} \right\rangle_t + \left\langle E_{kbvr} \right\rangle_t + \left\langle E_{kbp} \right\rangle_t \cong$$

$$\pi \rho l D^4 \Omega^2 \left( \frac{1}{4} + \ln \frac{R_{en}}{D} \right) + \frac{\pi \rho \Omega^2 R_0^3 r_{b0}^2}{3} + \frac{\pi \rho R_0^9 \omega^2 a^2 \sin^2 \varphi_b}{3 r_{b0}^4} + 2\pi \rho R_0^5 a^2 \omega^2 \cong \tag{76}$$

$$\pi \rho l D^4 \Omega^2 \left( \frac{1}{4} + \ln \frac{R_{en}}{D} \right) + \frac{\pi \rho R_0^3 \Omega^2 r_{b0}^2}{3} + 2\pi \rho R_0^5 a^2 \omega^2$$

and b) outside of the core

$$\left\langle E_{kbve} \right\rangle_t = E_{kv} + \left\langle E_{kboe} \right\rangle_t + \left\langle E_{kbvo} \right\rangle_t + \left\langle E_{kbp} \right\rangle_t \cong$$

$$\pi \rho l D^4 \Omega^2 \left( \frac{1}{4} + \ln \frac{R_{en}}{D} \right) + \frac{\pi \rho \Omega^2 R_0^3 r_{b0}^2}{3} + \frac{\pi \rho R_0^9 \omega^2 a^2 \sin^2 \varphi_b}{3 r_{b0}^4} + 2\pi \rho R_0^5 a^2 \omega^2 \cong \tag{77}$$

$$\pi \rho l D^4 \Omega^2 \left( \frac{1}{4} + \ln \frac{R_{en}}{D} \right) + \frac{\pi \rho R_0^3 D^4 \Omega^2}{3 r_{b0}^2} + 2\pi \rho R_0^5 a^2 \omega^2.$$

Under the conditions $\rho \gg \rho_b$, $R_{en} \gg D$, and $R_0 a \omega \ll r_{b0} \Omega$, the expressions of the total kinetic energy are approximated by:

$$\langle E_{kbvi} \rangle_t \cong \pi \rho l D^4 \ln \frac{R_{en}}{D} \Omega^2 + \frac{\pi \rho R_0^3 \Omega^2 r_{b0}^2}{3} = \pi \rho \left( l D^4 \ln \frac{R_{en}}{D} + \frac{R_0^3 r_{b0}^2}{3} \right) \Omega^2,$$

$$\langle E_{kbve} \rangle_t \cong \pi \rho l D^4 \ln \frac{R_{en}}{D} \Omega^2 + \frac{\pi \rho R_0^3 D^4 \Omega^2}{3 r_{b0}^2} = \pi \rho D^4 \left( l \ln \frac{R_{en}}{D} + \frac{R_0^3}{3 r_{b0}^2} \right) \Omega^2.$$

(78)

The potential energy of the system is the potential energy of the bubble (more generally, of the spherical body with density $\rho_b$) in the vortex field. The forces between the bubble and the vortex being attractive both in the core and outside the core, the potential energy is negative. We find the modulus of the potential energy by calculating the mechanical work of the vortex-bubble forces

$$E_{pvb} = -\int_\infty^{r_b} \vec{F}_{vb} d\vec{r} = \int_\infty^{r_b} F_{vb} dr, \; E_{pvbi} = \int_\infty^{D} F_{vbe} dr + \int_D^{r_b<D} F_{vbi} dr, \; E_{pvbe} = \int_\infty^{r_b>D} F_{vbe} dr. \tag{79}$$

We find the expressions for the internal and external forces using the radial accelerations given by Eqs. (19, 21) and neglecting the oscillations of the bubble ($R^3 \cong R_0^3$, $m_b \cong 2\pi R_0^3 \rho/3 = m_{b0}$)

$$F_{vbi} = m_b \dot{v}_{bi} = m_b C \Omega^2 r_b, \; F_{vbe} = m_b \dot{v}_{be} = m_b C \frac{\Omega^2 D^4}{r_b^3}. \tag{80}$$

Substituting, the force expressions given by Eqs. (80) in the expressions of the potential energy we obtain: a) in the core of the vortex ($r_b < D$)

$$E_{pvbi} = \int_\infty^D F_{vbe} dr + \int_D^{r_b<D} F_{vbi} dr = m_b C\Omega^2 D^4 \int_\infty^D \frac{dr}{r^3} + m_b C\Omega^2 \int_D^{r_b<D} r \, dr =$$

$$\left( -m_b C\Omega^2 D^4 \right) \frac{1}{2D^2} + m_b C\Omega^2 \left( \frac{r_b^2}{2} - \frac{D^2}{2} \right) = m_b C\Omega^2 \left( \frac{r_b^2}{2} - D^2 \right) < 0$$

(81)

and b) outside the vortex core ($r_b > D$)

$$E_{pvbe} = \int_\infty^{r_b>D} F_{vbe} dr = m_b C\Omega^2 D^4 \int_\infty^{r_b>D} \frac{dr}{r^3} = \frac{-m_b C\Omega^2 D^4}{2 r_b^2} < 0. \tag{82}$$

These averaged potential energies have the expression: a) in the core of the vortex ($r_b < D$)

$$\langle E_{pvbi} \rangle_t = \left\langle m_b C\Omega^2 \left( \frac{r_b^2}{2} - D^2 \right) \right\rangle_t \cong m_{b0} C\Omega^2 \left( \frac{r_{b0}^2}{2} - D^2 \right) \tag{83}$$

and b) outside the vortex core ($r_b > D$)

$$\langle E_{pvbe} \rangle_t = \left\langle \frac{-m_b C\Omega^2 D^4}{2 r_b^2} \right\rangle_t \cong \frac{-m_{b0} C\Omega^2 D^4}{2 r_{b0}^2}. \tag{84}$$

The time-averaged kinetic energies of the bubble in orbital motion $\langle E_{tbvi} \rangle_t$ are according to Eqs. (72, 73). The total energies averaged over time, neglecting the radial oscillation kinetic energies of the system, are: a) in the core of the vortex

$$\langle E_{tbvi} \rangle_t = \langle E_{kbvi} \rangle_t + \langle E_{pbvi} \rangle_t = E_{kvi} + \langle E_{kboi} \rangle_t + \langle E_{kbp} \rangle_t + \langle E_{pbvi} \rangle_t \cong$$

$$\pi \rho l D^4 \Omega^2 \left( \frac{1}{4} + \ln \frac{R_{en}}{D} \right) + \frac{m_{b0} C\Omega^2 r_{b0}^2}{2} + 3 m_{b0} R_0^2 a^2 \omega^2 + m_{b0} C\Omega^2 \left( \frac{r_{b0}^2}{2} - D^2 \right) =$$

$$\pi \rho l D^4 \Omega^2 \left( \frac{1}{4} + \ln \frac{R_{en}}{D} \right) + 3 m_{b0} R_0^2 a^2 \omega^2 + m_{b0} C\Omega^2 \left( r_{b0}^2 - D^2 \right).$$

(85)

and b) outside the vortex core ($r_b > D$)

$$\langle E_{tbve} \rangle_t = \langle E_{kbve} \rangle_t + \langle E_{pbve} \rangle_t = E_{kve} + \langle E_{kboe} \rangle_t + \langle E_{kbpe} \rangle_t + \langle E_{pbve} \rangle_t \cong$$
$$\pi \rho l D^4 \Omega^2 \left( \frac{1}{4} + \ln \frac{R_{en}}{D} \right) + \frac{m_{b0} C D^4 \Omega^2}{2 r_{b0}^2} + 3 m_{b0} R_0^2 a^2 \omega^2 - \frac{m_{b0} C \Omega^2 D^4}{2 r_{b0}^2} = \quad (86)$$
$$\pi \rho l D^4 \Omega^2 \left( \frac{1}{4} + \ln \frac{R_{en}}{D} \right) + 3 m_{b0} R_0^2 a^2 \omega^2.$$

We note that the total energy states $\langle E_{tbvi} \rangle_t$ are more stable than the energy states $\langle E_{tbve} \rangle_t$

$$\langle E_{tbvi} \rangle_t = \pi \rho l D^4 \Omega^2 \left( \frac{1}{4} + \ln \frac{R_{en}}{D} \right) + 3 m_{b0} R_0^2 a^2 \omega^2 + m_{b0} C \Omega^2 \left( r_{b0}^2 - D^2 \right) =$$
$$\langle E_{tbve} \rangle_t + m_{b0} C \Omega^2 \left( r_{b0}^2 - D^2 \right) < \langle E_{tbve} \rangle_t. \quad (87)$$

since, for $r_b < D$, the energy $m_{b0} C \Omega^2 \left( r_{b0}^2 - D^2 \right)$ is negative.

## 5. Comments, discussions and conclusions

We studied the vortex-bubble system with the intention of verifying whether it is a good model for an acoustic particle with spin (internal kinetic momentum) and electric acoustic charge (acoustic monopole).

The obtained results do not highlight, for this system, a magnetic acoustic field for the outer vortex. According to the work Vortexes as systems specific to the Acoustic World [3], we obtained that the inner vortex generates both an acoustic charge and an electric acoustic field as well as a magnetic acoustic field Eqs (50, 51) and in the outer core vortex only an electric acoustic field is generated analogous to the field of an electric dipole. Using the equations of motion of a bubble in a vortex from the paper [17] we obtained the equations of motion of the bubble in the vortex field (the expression of the acceleration of the bubble depending on the angular velocity, the radius and the velocity of the bubble). To solve the equations, we made the assumption that the speed of the bubble is proportional to the speed of the vortex.

We found the proportionality constant, $B$, depending on the densities of the fluid and the bubble by imposing the condition that the acceleration of the bubble is equal to the centrifugal acceleration (that is, the condition of stability on the orbit of the bubble) expressed as a function of the bubble's speed and radius. The obtained result can be interpreted as a change in the angular speed of the bubble ($\upsilon_{vi} = (B\Omega) r = \Omega_b r$) or a change in the radius of the bubble's orbit ($\upsilon_{vi} = (Br)\Omega = r_b \Omega$). We could not solve the situation in which both the radius and the angular velocity change. We calculated the total kinetic momentum of the system and the total energy for the orbital steady states. The bubble-vortex system has a spin-angular moment of the vortex and an orbital-angular moment of the bubble. We have not studied the rotational motion around the bubble axis determined by the fluid velocity gradient in the vortex and hence the bubble spin angular moment.

The total energy expression shows that the orbital states in the vortex core are more stable than those outside the vortex due to the negative potential energy. A restrictive condition, such as the quantification of the kinetic moment, would lead to the highlighting of two states, one with a lower total energy, for $r_b < D$, and one with a higher and less stable energy, for $r_b > D$, but with the same electrical acoustic charge (situation analogous to the existence of the electron and the muon [24] in the electromagnetic World). For the vortex-bubble system, we did not study the magnetic acoustic field generated by the acoustic charge of the bubble in orbital motion and the emission of acoustic radiation that determines the instability of the circular trajectory. The study of vortices highlights other vortex-bubble systems such as the vortex ring system [25] with an annular bubble trapped inside [26]. The annular vortex has the property of being in translational motion with velocity perpendicular to the plane of the ring and having zero angular momentum. One can also imagine and study a vortex ring with zero angular momentum that also rotates around the translational velocity axis. The counterpart of this acoustic particle would be the particle called neutrino [24]. The purpose of the theoretical investigation of these phenomena in the AW is to guide the study of a particle model with electric charge and internal angular momentum in the electromagnetic world, i.e. a "bubble" captured by a vortex in a superfluid.

We can have a better understanding of the phenomena related to the vortex-bubble system only after we study the dynamics of the system formed by vortex-bubble systems (clusters formed by $N$ vortex-bubble systems).

According to the properties of the vortex-bubble system studied in this paper we can conclude that this system is a good model for the acoustic particle with charge and internal angular momentum.